\documentstyle[preprint,aps]{revtex} 
\begin{document}

\draft

{\tighten
\preprint{\vbox{\hbox{WIS-96/38/Oct-PH}
                \hbox{hep-ph/9610540} }}

\title{Renormalization Group  Induced Neutrino 
Mass in Supersymmetry without R-parity}

\author{Enrico Nardi}

\footnotetext{ 
E-mail address: ftnardi@wicc.weizmann.ac.il 
}

\address{ 
\vbox{\vskip 0.truecm}
  Department of Particle Physics \\
  Weizmann Institute of Science, Rehovot 76100, Israel \\ }

\maketitle

\begin{abstract}%
We study supersymmetric models without $R$ parity and with universal 
soft supersymmetry breaking terms. We show that as a result  
of the renormalization group flow of the parameters, a misalignment between 
the directions in field space of the down-type Higgs vacuum expectation 
value $v_d$ and of the $\mu$ term is always generated. 
This misalignment induces 
a mixing between the neutrinos and the neutralinos, resulting in one massive 
neutrino. By means of a simple approximate  analytical expression, we study 
the dependence on the different parameters that contribute to the misalignment 
and to $m_\nu\,$. In large part of the parameter space this effect 
dominates over the standard one-loop contributions to $m_\nu\,$; 
we estimate  1 MeV 
${\mbox{${~\raise.15em\hbox{$<$}\kern-.85em\lower.35em\hbox{$\sim$}~}$}} 
m_\nu 
{\mbox{${~\raise.15em\hbox{$<$}\kern-.85em\lower.35em\hbox{$\sim$}~}$}} $ 
1 GeV.  Laboratory, cosmological and astrophysical 
constraints imply $m_\nu 
{\mbox{${~\raise.15em\hbox{$<$}\kern-.85em\lower.35em 
\hbox{$\sim$}~}$}} 100\,$eV. To be phenomenologically viable, 
these models must be supplemented with some additional 
mechanism to ensure approximate alignment and to suppress $m_\nu\,$.
\end{abstract}
} 

\newpage

\section{Introduction}

The field content of the Standard Model (SM) together with the
requirement of $SU(2)_L\times U(1)_Y$ gauge invariance, implies  
that at the renormalizable level the most general Lagrangian 
possesses additional accidental $U(1)$ symmetries. 
The $U(1)$ generators correspond to Baryon ($B$) and Lepton flavor 
($L_i$) charges. 
The conservation of $B\,$, $L_i\,$ and hence of total Lepton number 
($L=\sum_i L_i$) naturally explains nucleon stability as well as 
the non observation of $L$ and $L_i$ violating  transitions. 
This nice feature of the SM is lost in its 
Supersymmetric (SUSY) extensions: once the SM fields are promoted to 
superfields, additional gauge and Lorentz invariant  terms are allowed, 
which violate $B$, $L_i$ and $L$ at the renormalizable level. 
To forbid these dangerous terms, a parity quantum number 
$R=(-1)^{3B + L + 2S}$ ($S$ being the spin) is assigned to each 
component field, and invariance under $R$ transformation is imposed. 
Therefore, in SUSY frameworks  $B$ and $L$  quantum numbers 
are assigned  {\it ad hoc} to the superfields to reproduce 
the accidental symmetries of the SM, and ensure the absence of 
non observed transitions.
However, $R$ parity is by no means the only symmetry which 
allows for building  viable SUSY extensions of the SM \cite{Ross}. 
{}From a phenomenological point of view, the first priority is to 
ensure the absence of operators leading to fast nucleon decay, and  
in this respect other discrete symmetries can be more effective than $R$.
This is because $R$ parity only forbids dimension 4 $B$ and $L$ 
violating terms, but does not forbid dimension 5 operators which 
can still be dangerous, even when suppressed by factors as large as the 
Planck mass. Some interesting alternatives to $R$ parity exist,  
which  forbid dimension 4 and 5 $B$ violating terms but  
do not imply  the same for the $L$ non-conserving terms \cite{Ross}, and thus
imply a rather different phenomenology from models with   $R$-parity.   
Since a mild violation of $L$ can be phenomenologically tolerated, 
SUSY extensions of the SM with highly suppressed $B$ violation 
but without $R$ parity and without $L$ number,  
represent interesting alternatives to the Minimal Supersymmetric 
Standard Model (MSSM).
Two new types of Lagrangian  terms characterize this class of models: 
{\it (i)} renormalizable  interactions responsible for 
$L$ and $L_i$ violating transitions; 
{\it (ii)} superrenormalizable terms which mix the three lepton doublets  
with  the down-type Higgs. 
These  mixing terms are present because the four hypercharge $Y=-1$   
doublets transform in the same way under the full gauge group. 
They also imply that after the gauge symmetry  is spontaneously broken,  
all the color singlet fermion fields with the same electric charge are mixed 
(left and right handed charged leptons mix with Higgsinos and winos, 
neutrinos mix with neutral Higgsinos and with the zino). 
This situation can appear phenomenologically untenable, 
however the mixing  acquires a well defined physical meaning only 
when a physical basis for the various  fields is defined. 
We define the  down-type Higgs as the particular combination 
of the four $Y=-1$ doublets which acquires a non vanishing 
vacuum expectation value. If all the superrenormalizable terms 
in the Lagrangian are such that in this basis the remaining three 
combinations are decoupled from the Higgs, we can 
still assign to the fields a lepton number which  is violated only 
by  the  renormalizable interactions {\it (i)} while,   
at lowest  order, it is conserved by the mass terms and by the 
gauge interactions.
In the class of models  where  the soft terms responsible 
for SUSY  breaking are universal, the  conditions required to realize 
this scenario seem to be satisfied. Since minimal SUSY extensions 
of the SM generally belong to this class,  most of the 
literature on SUSY without $R$ parity concentrated in studying 
the effects of the interaction terms  {\it (i)} 
\cite{Aulakh,Hall-Suzuki,Lee,Dawson,Ross-Valle,Ellis,Santamaria-Valle,%
Brahm-Hall-Hsu,Romao-Valle,Joshi-Nowa,Campos,Vissani,BGNN,FBGNN,%
Barbieri-Masiero,Mohapatra1,Bar-Giu-Han,Dreiner-Ross,Mohapatra2,%
Hinchliffe-Kaeding,HKKK,Babu-Moha1,Bhatta-Ellis,Dimopoulos-Hall,%
Babu-Moha2,BGMT,Este-Tom,Mas-Riotto,Roy-Tata,Chun-Lukas}, 
while  less attention has been payed to the consequences 
of {\it (ii)} \cite{Aulakh,Hall-Suzuki,Lee,Dawson,Ross-Valle,Ellis,%
Santamaria-Valle,Brahm-Hall-Hsu,Romao-Valle,Joshi-Nowa,%
Campos,Vissani,BGNN,FBGNN}.   
However, even in the minimal models, universality of the soft 
terms holds only at some high energy scale where these terms 
are generated. The set of loop corrections induced by the terms  {\it (i)} 
imply deviations from universality for the low energy 
parameters, and this unavoidably results in the appearance 
of the terms {\it (ii)} which therefore have to be always included in 
the $R$-parity nonconserving superpotential. Moreover, since the scale 
where universality holds 
can be as large as the Planck scale, deviation from universality 
at low energy can be relevant and imply that the effects of 
the Renormalization Group (RG) induced superrenormalizable terms 
cannot be neglected. 

In Section II we first present a qualitative discussion, based on symmetry 
considerations, of the mass spectrum for the color singlet fermions.
We also review the conditions for the alignment in field space between 
the down-type Higgs vacuum expectation value and the $\mu$ term 
\cite{Hall-Suzuki,BGNN} which plays a crucial role in ensuring 
the suppression of neutrino masses. In Section III we derive a simple 
formula which parametrizes the RG induced misalignment, and we discuss the 
main dependence of this effect on the model parameters. 
The fermion mass spectrum is discussed 
quantitatively in Section IV, where the mass of the heaviest neutrino arising 
from the misalignment is estimated. Section V contains a brief review of 
the main laboratory, cosmological and astrophysical constraints on the 
neutrino mass, which can be translated into  constraints 
on the relevant parameters responsible for the misalignment. Our results 
are summarized in Section VI. 

In many aspects our analysis complements recent works that 
discuss the same effect \cite{Hempfling,Carlos-White,Nills-Polonsky}.
Refs. \cite{Hempfling} and  \cite{Nills-Polonsky} restrict their 
analysis to models in which $L$ violation enters only  through the 
bilinear terms {\it (ii)}. The renormalizable interactions {\it (i)} 
arise from the Yukawa terms only after the fields are rotated to a basis 
where the lepton doublets are decoupled from the Higgs. Therefore 
the form of these terms is not general, but is determined 
by the corresponding Yukawa couplings with a proportionality factor   
accounting for the field rotation.  Issues analogous to the ones 
studied here are also addressed in Section IV of \cite{Carlos-White}, 
where some results corresponding to specific choices of the parameters are 
presented. A brief discussion of these effects 
is also given in Section VI of \cite{Vissani}.   
Most of the results presented in these studies are 
given in numerical or graphical forms, which render difficult 
to appreciate the details of the physics involved. In the absence of     
analytical results it also appears awkward the task of taking properly 
into account these effects in future studies of SUSY models 
without $R$-parity. 
In the present work we study the  general $R$-parity violating case by 
including all the terms consistent with the SM gauge symmetry and with 
$B$ conservation.
In contrast to previous works, our approach is essentially analytical. 
We give a simple basis independent expression for the RG induced misalignment 
which highlights its physical meaning. We present an   
analytical formula for the neutrino mass that shows explicitly 
the main effects involved, and makes it easy to appreciate the  
various interrelations between the different parameters of the model. 
All our main results are summarized in a few 
simple expressions that can be easily used for investigating further 
this class of models.

\section{Alignment}
In this section we examine the qualitative features of  the  fermion 
mass spectrum which can be expected in SUSY models without $R$ parity. 
The SUSY extension of the 
SM contains eight  color-singlets chiral multiplets,   
corresponding to the up-type Higgs field,  
three right-handed leptons, three left-handed leptons 
 and the down-type Higgs doublets.  
Under the electroweak gauge group 
$ SU(2)_L \times U(1)_Y $ their quantum number assignments are 
\begin{eqnarray}\label{assignments}
\hat   H_u &\sim&\ (\,2,\, 1\,) \,, \cr
\hat {\bar \ell_i} \ \  &\sim&\ (\,1,\,2\,)\,, 
\qquad\qquad\qquad (i=1,2,3)   \cr
\hat H_\alpha &\sim&\ (\,2,\,-1\,)\,, \qquad\qquad (\alpha = 0,1,2,3)\,. 
\end{eqnarray}
Here  $\hat H_\alpha$ denotes collectively the supermultiplets containing 
the down-type Higgs and left-handed lepton doublets, which in the  MSSM are 
distinguished  by different R-parity assignments.  If R-parity is not 
imposed, the  gauge interactions posses a global $SU(4)$ symmetry 
corresponding to rotations of the four $\hat H_\alpha$ 
superfields~\cite{Nills-Polonsky}.
However, other  terms are generally present which select some preferred 
directions in $SU(4)$ field space.
The relevant terms that break the symmetry in the fermion sector are:
   
\vspace{-8pt}
\begin{itemize}\itemsep=-4pt
\item[($a$)]
The bilinear superpotential term 
\begin{equation} \label{bilinear}
  \mu_\alpha \hat H_\alpha \hat  H_u \,,
\end{equation}
provides a mass for the fermionic component of one 
combination of the $\hat H_\alpha$ doublets 
(the Higgsino).  The symmetry is broken  
down to $SU(3)$  acting  on the three combinations orthogonal 
to the Higgsino. 
\item[($b$)]
The vacuum expectation values  (vevs) 
\begin{equation} \label{vevs}
\langle{H_\alpha}\rangle =v_\alpha\,, 
\end{equation}
which contribute to the spontaneous breaking of the electroweak 
(EW) symmetry, induce a mixing between the neutral members of 
the $\tilde  H_\alpha$ fermion doublets and the neutral gauginos, 
thus breaking the symmetry down to $SU(2)$.  
A second combination of  the neutral members in $\tilde  H_\alpha$ 
acquires a mass in this way. Since the vector $v_\alpha$ 
fixes a direction only for the neutral  fields, 
no additional charged fermion becomes massive at this stage. 
\item[($c$)]
{}Finally, the following trilinear terms in the superpotential  
break the symmetry completely: 
\begin{equation} \label{trilinear}
 \lambda_{\alpha\beta k} \> \hat H_\alpha \, 
\hat H_\beta \,  \hat {\bar \ell_k} + 
 \lambda^\prime_{\alpha   j  k} \> 
\hat H_\alpha  \,\hat Q_j  \,  \hat {\bar d_k}\,.
\end{equation}
Here $\hat Q_j $ and $ \hat {\bar d_k} $ denote the quark doublet 
and down-quark singlet superfields, and 
$ \lambda_{\alpha\beta k} = -  \lambda_{\beta \alpha k}\>$  
due to the antisymmetry in the $SU(2)$ indices. 
{}For the charged fields the breaking is induced at tree level 
by the $ \lambda_{\alpha\beta k}$ couplings,   
which generate three new vectors in $SU(4)$ space 
corresponding to the  mass terms 
$(m^\ell)_{\alpha k} =  \lambda_{\alpha\beta k}\, v_\beta $, $(k=1,2,3)$. 
In the neutral sector the residual $SU(2)$ symmetry is 
broken only at the loop level, through quark-squark and 
lepton-slepton loop diagrams which generate the  mass 
terms~\cite{FBGNN,Dimopoulos-Hall,Babu-Moha2,BGMT,Este-Tom,%
Mas-Riotto,Roy-Tata,Chun-Lukas}
\begin{equation} \label{mnu}
m^\nu_{\alpha \beta} \simeq 
 {3\,  \lambda^\prime_{\alpha i j}  
\lambda^\prime_{\beta l k} \over 8 \pi^2}\,
{(m^d)_{i k}(\tilde  M^{d^{\scriptstyle \,2}}_{LR})_{j l}\over\tilde  m^2}\,  
+ \, { \lambda_{\alpha\gamma j}  \lambda_{\beta \sigma k} \over 8 \pi^2}\,
{(m^\ell)_{\gamma k} 
(\tilde  M^{\ell^{\scriptstyle \,2}}_{LR})_{j\sigma}\over\tilde  m^2}\,.
\end{equation}
Here $m^d$  is the $d$--quark mass matrix which 
arises at tree level from the second term in 
(\ref{trilinear}),   
$\tilde  M^{d^{\scriptstyle \,2}}_{LR}$ is the left--right sector in the 
$\tilde  d$-squark mass-squared matrix, 
$\tilde M^{\ell^{\scriptstyle \,2}}_{LR}$
is the left--right sector in the mass-squared matrix for
the charged $\tilde  \ell_j$--$H_\sigma$ scalars, $\tilde  m$ 
represents a slepton or  squark mass, and the expression  
holds at leading order in $\tilde M^2_{LR} / \tilde m^2\,$.
\end{itemize} \vspace{-8pt}

The qualitative features of the fermion mass spectrum for the fields  
in (\ref{assignments}) arising from the pattern ($a$)-($c$) are the 
following:
\vspace{-8pt}
\begin{itemize}\itemsep=-4pt
\item[({1})]
Only one combination of the charged $\tilde H_\alpha$ 
acquires a large mass of  order  $\mu = (\mu_\alpha\mu_\alpha)^{1/2}$
(or of the order of the EW breaking scale) while the remaining three charged 
fermions get masses proportional to (arbitrarily small) Yukawa couplings.  
\item[({2})]
Two neutral combinations of the  $\tilde H_\alpha$ doublets 
acquire small masses only at the loop level ($c$), while   
other two get large tree level masses as a consequence of ($a$) and ($b$).  
\end{itemize} \vspace{-8pt}

Since three neutral fermions (neutrinos) 
are known to be very light, 
(2) represents a major challenge for reconciling  this scenario with
our experimental knowledge of the fermion mass spectrum.  
The qualitative features of the predicted spectrum 
can be reconciled with the observations if  
some mechanism   ensures  that 
\begin{equation}\label{vmu}
v_\alpha \propto \mu_\alpha\,.
\end{equation}
If this relation is satisfied,  no new direction is 
singled out by the vevs $v_\alpha\,$,   
$SU(3)$ still remains a good symmetry  after  ($b$) and  
three neutral fermions acquire their mass  only 
through the loop effects  ($c$).  
As we will see below, in general 
(\ref{vmu}) cannot be  enforced as an exact (low-energy) relation.  
However, since  the tree level mass which is induced  
at stage ($b$) is proportional to the amount 
of $SU(3)$ breaking, an {\it approximate} alignment
between $v_\alpha$ and $\mu_\alpha$ can be sufficient to avoid 
conflicts with the limits on neutrino masses. 
 
The conditions for  $v_\alpha$ and $\mu_\alpha$ alignment were studied 
in \cite{Hall-Suzuki,BGNN}. The direction of $v_\alpha$ is determined by 
the minimum equations for the scalar potential, which depend on the soft 
SUSY breaking terms   
\begin{equation}\label{soft}
B_\alpha H_\alpha  
  H_u\,, \qquad\qquad 
 {\tilde m^2_{{\scriptscriptstyle H}_\alpha 
{\scriptscriptstyle H}_\beta}}\, H_\alpha H_\beta\,, 
\end{equation}
and on $\mu_\alpha\,$. Terms proportional to $ \lambda_{\alpha \beta j}$ and
 $ \lambda^\prime_{\alpha i j}\,$ as well as   the 
soft SUSY breaking trilinear $A$ terms which also carry $SU(4)$ indices, 
always involve a charged field and hence at lowest order do not 
contribute to determine  $v_\alpha\,$. 
Relation (\ref{vmu}) holds if the following two conditions are 
satisfied~\cite{BGNN}:
 \vspace{-8pt}
\begin{itemize}\itemsep=-4pt
\item[$(A)$]\qquad 
$\mu_\alpha$ is an eigenvector of ${\tilde m^2_{{\scriptscriptstyle H}_\alpha 
{\scriptscriptstyle H}_\beta}}\,$: \  \  
$ {\tilde m^2_{{\scriptscriptstyle H}_\alpha 
{\scriptscriptstyle H}_\beta}} \, \mu_\beta = {\tilde m^2} \,\mu_\alpha\,;$
\item[$(B)$]\qquad
$B_\alpha$  is proportional to $\mu_\alpha\,$: 
\ \ $B_\alpha  = B \, \mu_\alpha$.
\end{itemize} \vspace{-8pt}
To show this, let us   rotate the $H_\alpha$ fields to 
the basis $(H_\parallel,H_\perp)$ 
where $H_\parallel = \mu_\alpha  H_\alpha/ \mu $  
and  $H_\perp$ denotes the three  combinations orthogonal to $H_\parallel\,$. 
According to $(A)$, in this basis 
$\tilde m^2_{H_\parallel H_\beta} = \tilde m^2\> \delta_{\parallel\,\beta}\>$ 
$(\beta = \,\parallel,\perp)$ 
while $(B)$ implies that,  like   $\mu_\alpha\,$, also $B_\alpha\,$
has the only non-vanishing component along $H_\parallel\,$.
Then the solution of the minimum equations corresponds to 
$\langle{H_\perp}\rangle =0\,$.
The vector $v_\alpha$ is thus aligned with $\mu_\alpha\,$: 
$v_d\equiv(v_\alpha v_\alpha)^{1/2} = \langle{H_\parallel}\rangle
 = \mu_\alpha v_\alpha/ \mu\,$.

\section{Renormalization group Induced Misalignment}

In models where SUSY breaking is induced by universal 
soft breaking terms, both conditions $(A)$ and $(B)$  
of the previous section hold. 
However, these conditions are {\it exactly} satisfied only 
at the scale $\Lambda_U$  where the universal terms are induced. 
After defining  $B_\alpha \equiv B_{\alpha\beta}\, \mu_\beta\,$, 
the universality
conditions  at $\Lambda_U$ read
\begin{eqnarray}\label{scale}
({\tilde m^2_{{\scriptscriptstyle H}_\alpha 
{\scriptscriptstyle H}_\beta}})_{\Lambda_U} &=&   
{\tilde m_U^2}\,\, \delta_{\alpha\beta} \cr 
( B_{\alpha\beta} )_{\Lambda_U}  &=&   B_U \,\, \delta_{\alpha\beta}\,. 
\end{eqnarray}
As a result of the RG running of the 
parameters,   these relations  
become only approximate  at low energy,  and a misalignment between 
$\mu_\alpha$ and $v_\alpha$ is generated. 
The deviations 
from conditions $(A)$ and $(B)$ at a generic energy scale 
can be parametrized as
\begin{eqnarray}\label{DeltaBm}
\left( {1\over{\tilde m^2}}{\tilde m^2_{{\scriptscriptstyle H}_\alpha 
{\scriptscriptstyle H}_\beta}} - 
\delta_{\alpha\beta} \right)\, \mu_\beta  &=& 
\Delta_{\alpha\beta}^{\tilde m}  \,  \mu_\beta  \cr
\left(\,\, {1\over B} \>\, B_{\alpha\beta}  \,\, - 
\,\,\delta_{\alpha\beta} \, \right) \, \mu_\beta   
&=&  \Delta_{\alpha\beta}^B \,  \mu_\beta  
\end{eqnarray}
where ${\tilde m^2} \simeq  ({\rm det}\> 
{\tilde m^2_{{\scriptscriptstyle H}_\alpha 
{\scriptscriptstyle H}_\beta}})^{1/ 4}\,$ and $B= 
(B_\alpha B_\alpha)^{1/ 2}$. 
To estimate  the misalignment induced by the RG running, we write 
the result of the minimization of the 
(low energy) scalar potential as 
\begin{equation}\label{valpha}
v_\alpha = {v_d\over\hat\mu}\,
 [\,\delta_{\alpha\beta} + \Delta_{\alpha\beta}\,]\,\,\mu_\beta\,,  
\end{equation}
where $\Delta_{\alpha\beta} = \Delta_{\alpha\beta}^{\tilde m}  + 
\Delta_{\alpha\beta}^B$  accounts for the
misalignment induced by  violations of conditions $(A)$ and $(B)$, and 
the normalization factor  $\hat\mu \simeq\mu$.

It is now convenient to introduce two unit vectors 
$e^v_\alpha$ and $e^\mu_\alpha$ with components $v_\alpha/v_d$ and 
$\mu_\alpha/\mu\,$.
The misalignment can be quantified by means of an  angle $\xi$  defined as 
\begin{equation}\label{sinxi}
\sin\xi = |\,  e^\mu \wedge e^v \,|  \,.
\end{equation}
By means of (\ref{valpha}) we obtain 
\begin{eqnarray}\label{mis}
\sin^2\xi &=&
 {1\over2 } \sum_{\alpha,\beta} \> 
\left[\,(e^\mu_\alpha\,\Delta_{\alpha\gamma} - 
e^\mu_\beta\,\Delta_{\beta\gamma})\,e^\mu_\gamma\,\right]^2 \nonumber \\* 
&=& \> e^\mu\cdot\Delta^2\cdot e^\mu - (e^\mu\cdot \Delta\cdot e^\mu)^2\,. 
\end{eqnarray}

We note that all factors proportional to 
$\delta_{\alpha\beta}$ in $\Delta_{\alpha\beta}$  
cancel in (\ref{mis}).  
Therefore, in computing $\sin\xi$   it is sufficient to retain 
only the terms which carry non trivial $SU(4)$ indices. 
In particular,  
the contributions proportional to the up-quarks  
Yukawa couplings and to the gauge couplings can be dropped off,
and only terms involving the couplings in (\ref{trilinear}) 
and the corresponding trilinear soft SUSY breaking $A$ terms 
need to be kept.  
An approximate analytical expression for $\Delta_{\alpha\beta}\,$, 
obtained by  assuming constant coefficients 
and by integrating the RG equations in one step, is derived in the Appendix. 
Proceeding in this way, it is possible to single out 
the main effects which generate misalignment and to keep track
of the various interrelations among different parameters. 
Besides universality,   in the following
we will also assume  that at $\Lambda_U$ 
the trilinear soft SUSY breaking $A$ terms are proportional 
to the corresponding couplings in  (\ref{trilinear}). 
This assumption is made only for reasons of simplicity, since it allows 
factoring out the overall scale $A_U$ of the soft-breaking terms
and this gives simpler expressions. 
However,  the results of the analysis do not depend on this assumption, 
and it is straightforward to replace terms like  
$ A_U \lambda_{\alpha\beta i}$ with the more general  parameters 
$ A_{\alpha\beta i}\, $.
At the EW scale  $\sim m_Z\,$,  the RG induced misalignment  
matrix reads 
\begin{equation}\label{Dab}
\Delta_{\alpha\beta} =  { t_U\over8 \pi^2} 
 \left(3+ {A^2_U\over{\tilde m_U^2}} + {A_U\over B_U}\right)\, 
( \lambda_{\alpha \gamma i}\lambda_{\beta  \gamma i} + 
3 \lambda^\prime_{\alpha ij}   \lambda^\prime_{\beta  ij})
\,, 
\end{equation}
where  $t_U = \log{M_Z/ \Lambda_U}\,$,    $A_U\,$, $B_U$ and ${\tilde m_U^2}$  
are the soft SUSY breaking parameters at $\Lambda_U\,$,  
and  only terms inducing  $SU(4)$ rotations are displayed. 
In (\ref{Dab}) 
the term proportional to $A_U/B_U$ comes from the running of 
$B_{\alpha\beta}$, 
while all the others originate from  ${\tilde m^2}_{\alpha\beta}$.
We learn   the following :

\vspace{-8pt}
\begin{itemize}\itemsep=-4pt
\item[({\it a})]
The RG induced $v_\alpha$--$\mu_\alpha$ misalignment originates mainly from 
the running of the the soft-breaking scalar masses. 
{}For $A_U \sim B_U \sim \tilde m_U$ 
this effect dominates by a factor of $\sim 4$.
Only if $B_U\ll A_U\ll \tilde m_U$ the misalignment is dominantly induced 
by the evolution of the $B$ terms. 
\item[({\it b})]
Apart from fine tuned cancelations and as long as $L$ is a broken symmetry, 
there is no limit for the soft-breaking 
terms in which  alignment can be recovered. 
\item[({\it c})]
If $A_U \ll B_U,\,\tilde m_U$, the misalignment is independent 
of the initial  values of the soft-breaking parameters. 
\item[({\it d})]
Since  $SU(4)$ rotations in the evolution 
of $B_{\alpha\beta} $ are induced only by terms 
proportional to $A_U$ (see the Appendix) if $B_U=0$ 
the third term in the first parenthesis in (\ref{Dab}) is unity. If 
$A_U=B_U=0\,$ the $B$ term does not contribute 
(at this order) to the misalignment. 
\end{itemize} \vspace{-8pt}

{}From (\ref{mis}) we see that $\sin\xi$ is 
a basis independent physical parameter. It 
is convenient to give its explicit expression in a specific basis. 
We define the  down-type Higgs field $H_d$  at the EW scale
by the condition $\langle{H_d}\rangle = v_d\,$ (that is   
$H_d= e^v_\alpha H_\alpha\,$) and we choose the basis 
$\{\hat H_d,\hat L_i\}$ where $\hat L_i$ are three states orthogonal 
to $\hat  H_d$. In this basis  $e^v_\alpha = \delta_{0\alpha}\,$, while  
$h^d_{ij} = \lambda^\prime_{\alpha j k} e^v_\alpha \,$   
and $h^\ell_{\beta  k} = \lambda_{\alpha \beta  k} e^v_\alpha \,$ 
(with $h^\ell_{0 k}=h^\ell_{\alpha  k} e_\alpha^v = 0 $ for the antisymmetry 
of the $\lambda$ couplings) are the $H_d$  Yukawa couplings to the fermions.
After inserting (\ref{Dab}) in  (\ref{mis}) and using  
$e^\mu_\alpha \simeq (\delta_{\alpha\beta} - 
\Delta_{\alpha\beta}) e^v_\beta  \simeq \delta_{0\alpha}$ 
we obtain 
\begin{eqnarray}\label{sinxib}
\sin^2\xi & \simeq  & \sum_i\, \Delta_{i0}\, \Delta_{i0} \cr &=&
\left({ t_U\over8 \pi^2} \right)^2 \> 
\left(3+ {A^2_U\over{\tilde m_U^2}} + {A_U\over B_U}\right)^2 \, \sum_i\> 
\left( h^\ell_{j k}\lambda_{i j k} + 
3\, h^d_{jk} \lambda^\prime_{ ijk}\right)^2\,, 
\end{eqnarray}
where it is understood that $\lambda_{i j k}$ and $\lambda^\prime_{i j k}$ 
are now the couplings  (\ref{trilinear})  rotated to the 
$\{\hat H_d,\hat L_i\}$  
basis.\footnote{For simplicity we do not distinguish between 
the couplings at the EW scale and at $\Lambda_U$.    
For our approximate solutions of the RG equations the difference is 
formally of higher order.}
{}From (\ref{sinxib}) we learn the following: 

\vspace{-8pt}
\begin{itemize}\itemsep=-4pt
\item[({\it e})]
To generate misalignment is enough to have at least one of the $L$ violating  
 $\lambda$ or  $\lambda^\prime$  couplings  
(or one of the corresponding $A$ terms, 
if the  assumption of proportionality is dropped)  non-vanishing. 
\item[({\it f})]
Assuming no particular suppression of the 
R-parity violating $b$-quark couplings $ \lambda^\prime_{i33}$ 
with respect to the 
couplings involving the first two families, the dominant contribution 
to the misalignment is proportional to the $b$-quark  
Yukawa coupling $h^d_{33}$ .  
\item[({\it g})] 
Since only the leptons and $d$-quarks couplings appear in (\ref{sinxib}), 
the misalignment  depends strongly also on the value of $\tan\beta = v_u/v_d$
(where $v_u=\langle{H_u}\rangle $). 
{}For the leading contributions we obtain 
\begin{equation}\label{num}
\sin^2\xi  \simeq  
\left[{3\, t_U\, m_b\over 8 \pi^2\, v} \right]^2 \, 
\left(3+ {A^2_U\over {\tilde m_U^2}} + {A_U\over B_U}\right)^2\, 
\> \left(\sum_i\>  \lambda^\prime_{ i33} \lambda^\prime_{ i33}\right) 
\> (1+\tan^2\beta)\,, 
\end{equation}
where $v=(v_d^2+v_u^2)^{1/2}\simeq 246\,$GeV. 
If in addition we assume that the $L$ violating couplings are proportional to 
the corresponding Yukawa couplings 
(as is the case in models based on horizontal symmetries 
\cite{BGNN,FBGNN}) then  $\sin^2\xi \sim \tan^4\beta\,$. 
\end{itemize} \vspace{-8pt}

\section{Fermion mass spectrum}

In this section we investigate the consequences 
of the RG induced misalignment 
on the fermion mass spectrum. Numerically, the  factor in square brackets in 
(\ref{num}) is at most of order $ 1\%\,$ resulting in $\sin\xi \ll 1$ and 
approximate low energy alignment. As we will see, this implies that $L$ 
violation in the mass terms is small, and that a distinction between 
`leptons' and charginos and neutralinos is still a meaningful one. 
We define the `right handed  leptons'  as the mass eigenstates having 
as main components the three $SU(2)$ singlets $\bar\ell_i$. Their mass 
partners are the  `left handed charged leptons' which are dominantly 
combinations of just the $Y=-1$ doublets.  
Their neutral $SU(2)$ partners constitute the main components of the 
neutrinos, while the remaining mass eigenstates are the charginos and the 
neutralinos. 

The mass matrix for the charged fermions $M_{\rm c}$ is $5\times5$, with
rows corresponding to $\{\tilde W^-,\tilde H_\alpha^-\}$, and columns to
$\{\tilde W^+,\tilde H_u^+,\bar\ell_k^+\}\,$:
\begin{equation}\label{Mc}
M_{\rm c}=\pmatrix{
M_2&{g\over\sqrt2}v_u&0_{1\times3}\cr
{g\over\sqrt2}v_\alpha&\mu_\alpha& \lambda_{\alpha\beta k}v_\beta\cr}\,.
\end{equation}
Here $0_{1\times3}$  denotes a zero $1\times3$ block and $M_2$ is 
the $SU(2)_L$ gauginos Majorana mass.
In the  $\{\tilde W^-,\tilde   H_d, L_i^-\}\,$ basis 
(in which we denote particles and superparticles according to the 
usual convention) this becomes  
\begin{equation}\label{Mcp}
M_{\rm c}^{\prime}=
\pmatrix{M_2&\sqrt2 m_W \sin\beta &0_{1\times3}\cr
\sqrt2 m_W \cos\beta &\mu \cos\xi &0_{1\times3}\cr 
0_{3\times1}&\mu\, (e^v\wedge e^\mu)_i &h^\ell_{i j} v_d\cr}\,, 
\end{equation}
where $m_W=gv/2$.  
As expected the charged lepton masses originate from the Yukawa
couplings to $H_d$, and their mixings with the charginos, induced by the 
$(3\>2)$ block,  is suppressed at least as $\sin\xi\,$. 
(If $\sin\xi\,$ is not too small, this mixing could 
still give rise to interesting processes like 
$Z\to \tilde W^+ L^-_i\,$, $L^+_iL^-_j\,$, 
$\bar \ell^+_i\bar \ell^-_j$ ($i\neq j$), etc.)  

The full neutralino mass matrix is $7\times7$. In the 
basis with rows and columns corresponding to
$\{\tilde B,\tilde W_3, \tilde H_u^0,\tilde H_\alpha^0\}$ it reads    
\begin{equation}\label{Mn}
M_{\rm n}=
\pmatrix{
 M_1 & 0 & {g\tan\theta_W\over 2}\> v_u & -{g \tan\theta_W\over 2} \> 
v_\alpha \cr 
 0   & M_2 &         -{g\over 2}\> v_u      &   {g\over 2}\> 
v_\alpha               \cr
{g \tan\theta_W\over 2} \> v_u  &        -{g\over 2}\> v_u     & 0 &
 -\mu_\alpha \cr
  -{g \tan\theta_W\over 2} \> v_\alpha &  {g\over 2}\> v_\alpha   
&-\mu_\alpha&0_{4\times4}\cr}\,. 
\end{equation}
Here $M_1$ is the $U(1)_Y$ gaugino mass and $\theta_W$ is the weak mixing 
angle.   In the $\{\tilde   H_d, L_i^-\}\,$ basis 
$v_\alpha = (v_d,0,0,0)$ and $\mu_\alpha \simeq (\mu,-\Delta_{i0} \mu)$.  
$M_{\rm n}$ gives 5 massive states and two massless ones.
{}Four massive states  correspond to the neutralinos while the fifth  
one, a neutrino, corresponds to a combination of 
the neutral members in $L_i$.  
The mass of the neutrino is given by 
\begin{equation}\label{numass}
m_\nu \simeq {{\rm det}^\prime M_n \over \> 
{\rm det}^\prime M_n\Big|^{}_{\xi=0}}\,, 
\end{equation} 
where ${\rm det}^\prime$ denotes the product of the nonvanishing eigenvalues
of the respective mass matrices. We have 
\begin{eqnarray}\label{det}
{\rm det}^\prime M_n &=&\mu^2\, M_{\tilde \gamma}\, m_Z^2\cos^2 
\beta\> \sin^2\xi\,, \cr 
{\rm det}^\prime M_n\Big|^{}_{\xi=0} &=&  \mu \> M_{\tilde \gamma} \,    
 m_Z^2  \, \sin2\beta - \mu^2 \, M_1  \, M_2\,,   
\end{eqnarray}
where $m_Z=gv/(2\cos\theta_W)$ and 
$M_{\tilde \gamma} = M_1 \cos^2\theta_W + M_2 \sin^2 \theta_W$ is    
the photino mass term. 
The final expression for the mass of the neutrino reads
\begin{equation}\label{numassf}
m_\nu \simeq  \mu \left[\sin2\beta - 
{ \mu \, M_1  \, M_2 \over M_{\tilde \gamma}\, m_Z^2}\right]^{-1}
\left[{3\, t_U\, m_b\over 8 \pi^2\, v} \right]^2 \, 
\left(3+ {A^2_U\over{\tilde m_U^2}} + {A_U\over B_U}\right)^2\, 
\> \left(\sum_i\>  \lambda^\prime_{ i33} \lambda^\prime_{ i33}\right)\,. 
\end{equation}
{}From this expression we see that the numerical value of 
$m_\nu$ depends on several parameters, and this explains why it 
is not easy to derive any simple scaling behavior from a numerical 
study of these effects. 
However, the leading behaviors are as follows: 
if $\mu M_2 / m^2_Z  \gg \sin2\beta\,$ (which is more easily satisfied  
in the $\tan\beta \gg 1$ limit) then $m_\nu \propto m^2_Z/M_2$, practically  
independently of $\mu$ and $\tan\beta\,$. Therefore, as was noted 
in~\cite{Hempfling},
$m_\nu$ vanishes in the limit of very large SUSY breaking scales.
In the opposite limit, which is consistent only for moderate values of 
$\tan\beta$ 
(${\mbox{${~\raise.15em\hbox{$<$}\kern-.85em\lower.35em\hbox{$\sim$}~}$}} 5$) 
and for small values of  $M_2$ and $\mu\,$, 
$m_\nu$ is approximatively proportional to  $\mu\,\tan\beta$.
It is interesting to note that in the first equation  (\ref{det}) 
$\cos^2\beta\,$  cancels against the explicit $1+\tan^2\beta$ term in 
the misalignment parameter $\sin^2\xi$ (\ref{num}), leaving only a mild 
dependence on $\tan\beta$ in the final result. However, if   
the $R$-parity  violating couplings are proportional 
to the Yukawa couplings, an implicit $\tan^2\beta$ dependence
from the misalignment is still present in the last parenthesis in 
(\ref{numassf}). 
We also note that the first square bracket in (\ref{numassf})
cannot approach zero, since is bounded by the lower limits 
on the neutralino masses. Banning possible fine tunings,  
for natural values of the parameters we obtain that
$\mu$ divided by  the first bracket yields a 
dimensionful factor $\sim 10-100\,$GeV.\footnote{The lower limit 
can be pushed down to 1 GeV in a phenomenologically viable scenario 
in which  $\mu$ is very small resulting in two neutralinos not much 
heavier than a few GeV \cite{Feng-Polonsky-Thomas}}
The square of the second  bracket provides a suppressing 
factor in the range $10^{-3}$--$10^{-5}$ corresponding respectively to 
$\Lambda_U \sim m_{\rm Planck}$  and $\Lambda_U \sim 10^5\,$GeV, 
where the second value is typical of gauge mediated SUSY breaking scenarios 
\cite{DNeS,DNeNS,DDRT,HIY,Carone-Hall,DNS}. {}Finally, the square of the 
term containing the soft breaking parameters yields approximatively a 
one order of magnitude enhancing factor.

As a result, in the absence of further suppression from the 
$\lambda^\prime$ couplings, we would estimate 1 MeV 
${\mbox{${~\raise.15em\hbox{$<$}\kern-.85em\lower.35em\hbox{$\sim$}~}$}} 
m_\nu 
{\mbox{${~\raise.15em\hbox{$<$}\kern-.85em\lower.35em\hbox{$\sim$}~}$}} $ 
1 GeV. 
 
Before concluding this section,  
it is interesting to compare the RG induced effects on $m_\nu$  with the 
standard contributions to the neutrino mass matrix from one-loop
diagrams~\cite{FBGNN,Dimopoulos-Hall,Babu-Moha2,BGMT,%
Este-Tom,Mas-Riotto,Roy-Tata,Chun-Lukas}. 
{}From (\ref{mnu}) the corresponding  leading  term reads 
\begin{equation}\label{leadingloop}
m^\nu_{ij} \simeq {3\over 8 \pi^2} \> 
{m_b^2\over {\tilde m}^2}\> \left(A-\mu \tan\beta\right)\>  
 \lambda^\prime_{i33} \lambda^\prime_{j33}.
\end{equation} 
The misalignment yields the dominant contribution 
to the mass of the heaviest neutrino as long as 
\begin{equation}\label{tulimit}
t_U {\mbox{${~\raise.15em\hbox{$>$}\kern-.85em\lower.35em\hbox{$\sim$}~}$}} 
\left[ {8 \pi^2 \over 3} 
\> F(m_{\rm soft})\,\right]^{1/2}\,,
\end{equation}
where the dimensionless function $ F $ depends in a  complicated way on the 
various soft SUSY breaking parameters, as well as on $\mu\,$ and $\tan\beta$. 
{}For most values of $F\,$ (\ref{tulimit}) is  satisfied as long 
as $\Lambda_U > 10^5\,$GeV, and thus in general the induced misalignment 
gives the leading effect. 
This implies that  predictions for the neutrino masses 
in models without $R$-parity and with high energy alignment   
based  solely on an estimate of the loop contributions~\cite{Chun-Lukas}, 
should be modified to include this effect. 

On the other hand, $F$ is maximal when $A_U\ll \tilde m_U, B_U\,$ and in this 
limit we obtain 
$F {\mbox{${~\raise.15em\hbox{$<$}\kern-.85em\lower.35em\hbox{$\sim$}~}$}} 
(1/ g^2)\> (\mu M_2/\tilde m^2 )\tan\beta $. 
This situation is interesting since  $A_U=0$ can arise in  gauge mediated 
SUSY breaking models  \cite{DNeNS}. In this case, for values of the 
relevant parameters such that $F \sim (1/ g^2)\,$ and for  
$\Lambda_U 
{\mbox{${~\raise.15em\hbox{$<$}\kern-.85em\lower.35em\hbox{$\sim$}~}$}} 
10^6\,$GeV  the two effects yield 
contributions which can be comparable in magnitude. 
{}Finally, for small $A_U$ and  $\tan\beta$ rather large
 ($\mu M_2\tan\beta/\tilde m^2 
{\mbox{${~\raise.15em\hbox{$>$}\kern-.85em\lower.35em\hbox{$\sim$}~}$}} 
25\,$)  the one-loop contributions 
(\ref{leadingloop}) dominate over the misalignment effects up to  
$\Lambda_U = m_{\rm Planck}\,$, and thus determine the mass of the heaviest 
neutrino.

\section{Phenomenological Constraints}

As we have seen, in SUSY models without R-parity and without Lepton number 
the induced $v_\alpha$--$\mu_\alpha$ misalignment results in  one massive 
neutrino which, in the absence of suppression of the $L$ violating
trilinear couplings,  is naturally  in the range 
1~MeV~$
{\mbox{${~\raise.15em\hbox{$<$}\kern-.85em\lower.35em\hbox{$\sim$}~}$}} 
m_\nu 
{\mbox{${~\raise.15em\hbox{$<$}\kern-.85em\lower.35em
\hbox{$\sim$}~}$}}$~1~GeV. 
In this section we argue that laboratory and cosmological constraints 
imply that this window is excluded. In addition  the massive neutrino of these 
models is  very likely  stable on a cosmological time scale, and thus the 
cosmological limit $m_\nu 
{\mbox{${~\raise.15em\hbox{$<$}\kern-.85em\lower.35em\hbox{$\sim$}~}$}} 
100\,$eV applies. As a consequence, to render 
these models phenomenologically 
viable some mechanism to suppress the $L$-violating couplings in 
(\ref{numassf}) down to 
$\sum_i  \lambda^\prime_{ i33} \lambda^\prime_{ i33} 
{\mbox{${~\raise.15em\hbox{$<$}\kern-.85em\lower.35em\hbox{$\sim$}~}$}} 
10^{-4}$ -- $10^{-7}$ is called 
for.\footnote{Such a suppression can be easily accommodated in models 
for fermion masses based on Abelian horizontal symmetries \cite{LNS1,NS,LNS2}. 
The required suppression for the case when the soft SUSY breaking terms are 
not universal and the misalignment arises as a tree level effect was studied 
in \cite{BGNN,FBGNN}.  In contrast to that case which required horizontal 
charges for the $L_i$ doublets larger than $\sim 7\,$,  in the present 
scenario a sufficiently small neutrino mass is obtained with the more natural 
values $Q_H(L_i) \sim 0 - 3 $.}

The flavor composition of the  neutrino is determined by the relative 
rotation  in the $\{\tilde H_d,L_i\}$ basis
arising from the diagonalization of  the submatrix 
in (\ref{Mn}) containing $\Delta_{i0}$ (for the $\nu$'s) and of 
the Yukawa couplings matrix $h^\ell {h^\ell}^\dagger$ 
(for the left-handed leptons). This can be studied only by specifying further 
the model. We will avoid doing this,  and we will conservatively assume that 
our massive state is mainly $\nu_\tau$, so that the laboratory limit on  
the mass is $m_\nu < 23\,$MeV \cite{LEPnu}. 
With regards to the neutrino mixing 
angles, the discussion below is purely phenomenological and 
does not require any theoretical estimate. 

Cosmological considerations of the age and the present
energy density of the Universe provide constraints relating the mass
and lifetime of the neutrino. {}For masses in the range 
$100\,$ eV -- a few MeV the constraint reads  
$m^2_\nu\tau_\nu 
{\mbox{${~\raise.15em\hbox{$<$}\kern-.85em\lower.35em\hbox{$\sim$}~}$}} 
2\times10^{8}\,$ MeV$^2$ sec \cite{Harari} .
When  charged particles are present in the final state,  a  
stronger bound from the absence of  distortions in the cosmic 
microwave background radiation (CMBR) applies, $\tau_\nu 
{\mbox{${~\raise.15em\hbox{$<$}\kern-.85em\lower.35em\hbox{$\sim$}~}$}}
10^4\ \sec$.
Detailed studies of the effects of a massive $\nu_\tau$ during the 
nucleosynthesis era imply an even stronger limit 
$\tau_\nu 
{\mbox{${~\raise.15em\hbox{$<$}\kern-.85em\lower.35em\hbox{$\sim$}~}$}}
10^2\,$sec,  for masses in the range $0.5\,$MeV --  $35\,$MeV   
and  independently of the decay 
modes~\cite{Turner,Dolgov-Ira,Kawasaki,Dolgov,Madsen,Kimmo}.
{}For visible decay modes (final states containing $\gamma$ or $e^\pm$) also 
a lower bound on $\tau_\nu$  exists, $\tau_\nu 
{\mbox{${~\raise.15em\hbox{$>$}\kern-.85em\lower.35em\hbox{$\sim$}~}$}} 
10^8\,$sec. 
This bound follows from the limits on the gamma-ray
fluence around the time when the neutrinos from the Supernova 1987A
were detected  \cite{Sigl}. This set of constraints already suggests that
$m_\nu 
{\mbox{${~\raise.15em\hbox{$>$}\kern-.85em\lower.35em\hbox{$\sim$}~}$}} 
1\,$ MeV is very likely ruled out.  

In order to avoid  some (or all) of these constraints, the massive $\nu$ 
should decay fast enough, and preferably into invisible final states.  
However, most likely the dominant $\nu$ decay mode is $\nu\to e^+e^-\nu_e$ 
which proceeds via $W$--mediated tree level diagrams.
All other decay modes, as $\nu\to 3\nu_\ell$ or $\nu\to \gamma \nu_\ell$  
($\ell=e,\mu$) 
are flavor changing neutral current (FCNC) processes, and are 
more suppressed. In particular, the invisible decay mode into three neutrinos 
involves the {}FCNC $Z \bar \nu \nu_\ell$ vertex, which is quadratic 
in the neutrino mixing  with the isotriplet neutralino  $\tilde W_3$,
and hence very small. 
{}For the leading decay mode  the lifetime is 
\begin{equation}\label{nulife} 
\tau_\nu = \left({ m_\mu\over m_\nu}\right)^5 {\tau_\mu 
\over |U_{1\nu}|^2} \simeq {2.8 \times 10^4 \over |U_{1\nu}|^2 } 
\left({ 1 MeV \over m_\nu}\right)^5\, {\rm sec}\,,  
\end{equation}
where $\tau_\mu \simeq 2.2\times 10^{-6}$ has been used. 
On the other hand, peak and kink searches in 
$\pi\,$, $K$ and $\beta$ decays yield stringent upper limits 
on  $|U_{1\nu}|^2$. We have \cite{PDG96}
$|U_{e\nu}|^2 
{\mbox{${~\raise.15em\hbox{$<$}\kern-.85em\lower.35em\hbox{$\sim$}~}$}} 
5\times 10^{-6}\,$ ($1\times 10^{-4}\,$) 
for a mass of about 20 MeV (5 MeV), implying 
lifetimes in conflict with the nucleosynthesis (and CMBR) constraint.
{}For smaller masses the laboratory limits on the mixing parameters 
are less stringent.  
However, below about 3.5 MeV (1.5 MeV) the constraint from nucleosynthesis   
(and CMBR) is not satisfied even for maximal mixing.
{}For $m_\nu < 1\,$MeV only {}FCNC decay channels are open, implying that 
also the weaker mass-lifetime constraint from the age of the Universe
is not evaded. Therefore, we conclude that independently of the mass and 
mixing angles the $\nu$ decay rate is not fast enough to evade all the 
constraints, and the limit for cosmologically stable neutrinos holds. 

\section{Conclusions}
In this paper we have presented an analysis of SUSY models without $R$ parity
and without Lepton number. We have shown that even when universality  
of the soft SUSY breaking terms is assumed, at low energy  the vector  
$v_\alpha$ of the vevs of the hypercharge $-1$  doublets $H_\alpha\,$ 
is not aligned with the vector $\mu_\alpha$ of the generalized $\mu$-term 
$\mu_\alpha \hat H_\alpha \hat H_u\,$.
The misalignment is induced by the renormalization group flow of the 
parameters from the scale where the soft SUSY breaking terms are generated, 
down to low energy. We have derived a simple  analytical 
expression which describes the dependence of the misalignment on the relevant 
parameters. Our treatment is basis independent, and shows that this effect 
cannot be rotated away or neglected. In the basis where the fields are 
physical, the bilinear superpotential terms 
$\mu_i \hat L_i \hat H_u\,$ which violate Lepton number by one unit, 
are always present. 
A major consequence of $v_\alpha$--$\mu_\alpha$ misalignment
is that one neutrino becomes massive,
and the mass induced in this way is generally larger than the contributions 
from one-loop diagrams. We have estimated that in the absence of additional 
suppression 1~MeV~$
{\mbox{${~\raise.15em\hbox{$<$}\kern-.85em\lower.35em\hbox{$\sim$}~}$}} 
m_\nu 
{\mbox{${~\raise.15em\hbox{$<$}\kern-.85em\lower.35em\hbox{$\sim$}~}$}} 
$~1~GeV. 
A brief analysis of  various laboratory, cosmological and 
astrophysical constraints  strongly suggests that this neutrino is 
cosmologically stable, and thus its mass must be  below 100 eV. 
This bound can be  translated into a constraint on the $R$-parity  
violating trilinear couplings 
$\sum_i  \lambda^\prime_{ i33} \lambda^\prime_{ i33} 
{\mbox{${~\raise.15em\hbox{$<$}\kern-.85em\lower.35em\hbox{$\sim$}~}$}} 
10^{-4}$ -- $10^{-7}$.  
We conclude that, to be  phenomenologically viable, SUSY models 
without $R$-parity must be supplemented 
with some mechanism (as for example a horizontal flavor symmetry) yielding a 
sufficient suppression of these couplings. 

\acknowledgements
I thank F. M. Borzumati, M. C. Gonzales-Garcia, Y. Grossman  and Y. Nir 
for several  helpful discussions. 

\appendix\section{}
In this Appendix we compute the misalignment matrix 
\begin{equation}
\Delta_{\alpha\beta} =  
\Delta_{\alpha\beta}^B + \Delta_{\alpha\beta}^{\tilde m}
\end{equation} 
induced by the RG evolution of the  soft SUSY breaking parameters  
${\tilde m^2_{{\scriptscriptstyle H}_\alpha 
{\scriptscriptstyle H}_\beta}}$ and $B_{\alpha\beta}\,$
from the high scale $\Lambda_U$ down to the EW scale. The running  
is controlled by 
the RG equations 
\begin{eqnarray}\label{RGE}
\phantom{\Bigg|} {d {\tilde m^2_{{\scriptscriptstyle H}_\alpha 
{\scriptscriptstyle H}_\beta}}\over dt} &=& 
{{\tilde m^2}\over 16\pi^2} G_{\alpha\beta}^{\tilde m^2}\,, \cr
\phantom{\Bigg|} {d B_{\alpha\beta}\over dt} &=& 
{A \over 16\pi^2} G_{\alpha\beta}^B\,,  
\end{eqnarray}
where we have factored out the overall scale ${\tilde m^2}$ and $A$ 
of the soft SUSY breaking masses and of the trilinear 
soft breaking terms.  The boundary conditions at $\Lambda_U\,$, where by 
assumption universality holds, are given in (\ref{scale}).
We solve (\ref{RGE}) in first approximation,  
neglecting the scale dependence of the coefficients. This gives 
\begin{eqnarray}\label{solutions} 
\phantom{\Bigg|} \Delta_{\alpha\beta}^{\tilde m^2} 
&\equiv& {1\over\tilde m^2_U}\, 
{\tilde m^2_{{\scriptscriptstyle H}_\alpha 
{\scriptscriptstyle H}_\beta}} -\delta_{\alpha\beta} 
\simeq  {t_U\over16 \pi^2} 
(G_{\alpha\beta}^{\tilde m^2})_{\Lambda_U}\,,  \cr
\phantom{\Bigg|}  \Delta_{\alpha\beta}^B &\equiv& {1\over B_U}\, 
B_{\alpha\beta} - \delta_{\alpha\beta} 
\simeq  {A_U\over B_U}\,{ t_U\over16 \pi^2} 
(G_{\alpha\beta}^B)_{\Lambda_U}\,,   
\end{eqnarray}
where  $t_U = \log{(M_Z/ \Lambda_U)}\,$. 
The SUSY RG equations including R-parity violation 
have been recently presented in a number of papers 
\cite{Carlos-White,Bar-Ber-Phil,Dreiner-Pois}. 
The equations for the soft SUSY breaking terms can be read off 
from \cite{Carlos-White}. {}For the running of 
$\tilde m^2_{\scriptscriptstyle H_\rho H_\sigma}$ in (\ref{RGE}) 
we have 
\begin{equation}\label{Gm} 
{\tilde m^2} G_{\alpha\beta}^{\tilde m^2} = 
C_{\ {\alpha\beta}}^{H\> {\rho\sigma}}  \> 
\tilde m^2_{\scriptscriptstyle H_\rho
 H_\sigma} 
+ C_{\ {\alpha\beta}}^{\bar \ell \>\> ij}\> 
\tilde m^2_{\bar\ell_i \bar\ell_j}
+ C_{\ {\alpha\beta}}^{Q \>\> ij}\>\tilde m^2_{\scriptscriptstyle Q_i Q_j}  
+ C_{\ {\alpha\beta}}^{\bar d \>\> ij}\>\tilde m^2_{\bar d_i \bar d_j}
+ A^2 C^A_{\, {\alpha\beta}} + C^{G} \, \delta_{\alpha\beta}
\end{equation} 
where   
\begin{eqnarray}\label{Ccoef}
C_{\ {\alpha\beta}}^{H\> {\rho\sigma}}  &=& 
    \lambda_{\alpha \gamma i}\lambda_{\rho \gamma i}\delta_{\beta\sigma} +
    \lambda_{\beta  \gamma i}\lambda_{\rho \gamma i}\delta_{\alpha\sigma} +
            2 \lambda_{\alpha \sigma i}\lambda_{\beta  \rho i} + 
            3( \lambda^\prime_{\alpha ij}   \lambda^\prime_{\rho ij} 
\delta_{\beta\sigma} + 
            \lambda^\prime_{\beta  ij}  \lambda^\prime_{\rho ij}
\delta_{\alpha\sigma}) \cr 
 C_{\ {\alpha\beta}}^{\bar \ell \>\> ij}  &=& 2\, \lambda_{\alpha \gamma i} 
                 \lambda_{\beta  \gamma j} \cr 
C_{\ {\alpha\beta}}^{Q \>\> ij} &=& 6\,  \lambda^\prime_{\alpha ik}  
\lambda^\prime_{\beta  jk} \cr 
C_{\ {\alpha\beta}}^{\bar d \>\> ij} &=& 6\,  \lambda^\prime_{\alpha ki}  
\lambda^\prime_{\beta  kj} \cr 
C^A_{\, {\alpha\beta}} &=& 2\,  (\lambda_{\alpha \gamma i} 
\lambda_{\beta  \gamma i} + 
        3\, \lambda^\prime_{\alpha ij}  \lambda^\prime_{\beta  ij}) \cr 
C^{G} &=& - \sum_{\sigma=\rm all} g^2_1 Y_\sigma {\tilde m^2}_{\sigma\sigma} 
- 2 g^2_1 M^2_1 - 
            6  g^2_2 M^2_2\,.
\end{eqnarray}   
The running of the $B$ term is determined by 
\begin{equation}
G_{\alpha\beta}^B = D_{\alpha\beta} + D\, \delta_{\alpha\beta} 
\end{equation} 
with 
\begin{eqnarray}\label{Dcoef}
\phantom{\Bigg|} 
D_{\alpha\beta} &=& 2\, 
\left( \lambda_{\alpha \gamma i}\lambda_{\beta \gamma i} + 
    3 \lambda^\prime_{\alpha ij}\lambda^{\prime}_{\beta ij}\right)  
\nonumber \\* 
\phantom{\Bigg|} 
D\ \  &=& 2\, \left(3 h^u_{ij}h^{u}_{ij} +
       g_1^2  {M_1\over A\,} +3g^2_2 {M_2\over A} \right)\,,  
\end{eqnarray}
where $h^u_{ij}$ are the up-quark Yukawa couplings.
Using the boundary conditions (\ref{scale})  together with 
$ (\tilde m^2_{\bar\ell_i \bar\ell_j})_{\Lambda_U} =
(\tilde m^2_{\scriptscriptstyle Q_i Q_j})_{\Lambda_U} =
(\tilde m^2_{\bar d_i \bar d_j})_{\Lambda_U} = 
 {\tilde m_U^2}\,\, \delta_{ij} $ we 
obtain\footnote{In models where SUSY breaking 
is communicated to the visible sector via gauge interactions, 
the  SUSY breaking terms are flavor-symmetric but not universal.
The corresponding modifications of  equations (\ref{A9}) are
straightforward, and do not affect our conclusions.} 
\begin{eqnarray}\label{A9}
(G_{\alpha\beta}^{\tilde m^2})_{\Lambda_U}
&=& \left(  6 + 2\,{A^2_U\over\tilde m^2_U }\,\right)\,  
( \lambda_{\alpha \gamma i}\lambda_{\beta  \gamma i} + 
 3\,    \lambda^\prime_{\alpha ik}  \lambda^\prime_{\beta  ik}) -  
\left[\> \sum_{\sigma=\rm all} g^2_1 Y_\sigma  + 2 g^2_1 
{M^2_U\over\tilde m^2_U } +
 6  g^2_2 {M^2_U\over\tilde m^2_U }\right] \,  
\delta_{\alpha\beta}\cr && \cr 
(G_{\alpha\beta}^B )_{\Lambda_U} &=&  
2\> ( \lambda_{\alpha \gamma i}\lambda_{\beta \gamma i} + 
      3 \lambda^\prime_{\alpha ij}\lambda^{\prime}_{\beta ij}) \, +\,  
    2\, \left[3 h^u_{ij}h^{u}_{i j}+
g_1^2  {M_U\over A_U\,} + 3g^2_2 {M_U\over A_U} \right] 
\delta_{\alpha\beta}\,, 
\end{eqnarray}
where $M_U$ is the universal gaugino mass. 
The terms in square brackets which are proportional 
to $\delta_{\alpha\beta}$ do not generate misalignment, and 
for our purposes can be dropped off. Inserting the relevant terms of 
equations (\ref{A9}) into (\ref{solutions}) gives the expression (\ref{Dab}) 
for the misalignment matrix $\Delta_{\alpha\beta}$.

{\tighten

}  

\end{document}